\documentclass[twocolumn,preprintnumbers,amsmath,amssymb,aps,prl]{revtex4}
\usepackage{graphicx}
\usepackage{epsfig}
\usepackage{multirow}
\usepackage{color}
\usepackage{setspace}

\begin{document}

\title{A highly efficient two level diamond based single photon source}

\author{D. A. Simpson}
\email{simd@unimelb.edu.au}
\author{E. Ampem-Lassen}
\author{B. C. Gibson}
\author{S. Trpkovski}
\author{F. M. Hossain}
\author{S. T. Huntington}
\author{A. D. Greentree}
\author{L. C. L. Hollenberg}
\author{S. Prawer}

\affiliation{Quantum Communications Victoria, School of Physics,
University of Melbourne}

\begin{abstract}
An unexplored diamond defect centre which is found to emit stable
single photons at a measured rate of 1.6 MHz at room temperature
is reported. The novel centre, identified in chemical vapour
deposition grown diamond crystals, exhibits a sharp zero phonon
line at 734 nm with a full width at half maximum of $\sim$ 4 nm.
The photon statistics confirm the center is a single emitter and
provides direct evidence of the first true two-level single
quantum system in diamond.
\end{abstract}


\maketitle
\begin{figure}[hbt]\centerline{\includegraphics[width=0.8\columnwidth,clip]
{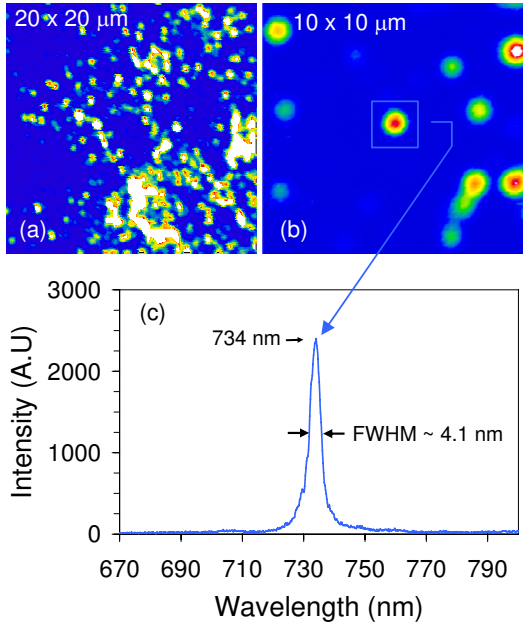}}\caption{(a) Fluorescence
intensity confocal map of emitting diamond crystals on a fused
silica substrate under 532 nm excitation. (b) High resolution
image of a 10$\times$10 $\mu$m region of the substrate. (c) The PL
spectrum from the emitting crystal identified in the centre of
(b).} \label{Fig:Confocal images and spectra}
\end{figure}

\begin{figure}[hbt]\centerline{\includegraphics[width=0.75\columnwidth,clip]
{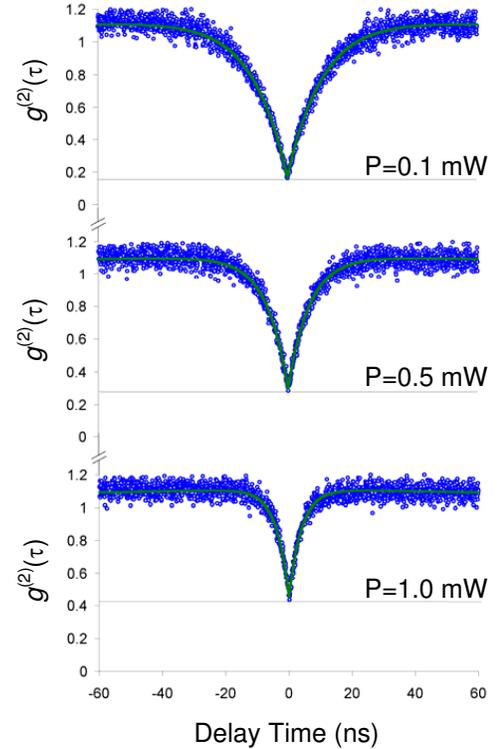}}\caption{Anti-bunching data from a
single 734 nm defect centre for excitation powers ranging from $P$
= 0.1-1.0 mW. The dots represent the measured coincidences whilst
the solid line was obtained from a theoretical fit as described in
the text. The pronounced dip in the coincidence counts at delay
time $t$ = 0 ns, illustrates the sub-Poissonian statistics of the
source.} \label{Fig:Anti-bunching vs pump power}
\end{figure}

\begin{figure}[hbt]\centerline{\includegraphics[width=0.9\columnwidth,clip]
{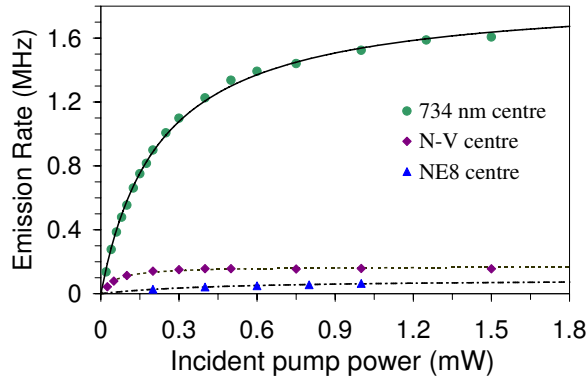}}\caption{Photon emission
rate for the 734 nm single emitting centre as a function of pump
power when excited at 532 nm. The emission rate is given by the
sum of the counts on the two APDs in the Hanbury Brown and Twiss
arrangement. The fit to the measured data is discussed in the
text. The measured emission rates for a single N-V and NE8 centre
are included in the figure as a comparison and were all measured
using the same experimental setup, with the exception of the NE8
data which was measured under 685 nm excitation.}
\label{Fig:Emission rate vs pump power}
\end{figure}

 Nanotechnology and photonics industries around the
world are currently undergoing a revolution as research into
individual quantum systems opens up new and exciting opportunities
in quantum information processing (QIP) \cite{1,20,21} and quantum
key distribution (QKD) \cite{2}. At the forefront of this
revolution is the development of light sources which emit single
particles of light on-demand, known as single photon sources
(SPSs). Single photon generation has been demonstrated with
several different technologies including quantum dots (QDs)
\cite{3}, single molecules \cite{4} and more recently carbon
nano-tubes \cite{5}; but perhaps the must promising and applicable
source of single photons for practical applications arise from
optical defects in diamond which can be easily accessed and
operate at room temperature. To date, of the over 500 known defect
centres in diamond only three have been identified as single
photon (SP) emitters \cite{6,7,8,9,10}. However, what sets these
color centres apart from the other technologies is their
uncompromised photo-stability at room temperature.  Until now, the
SP emission rates achieved in diamond based SPSs have been limited
to $\sim$ 300 kHz due to the presence of meta-stable/shelving
states which act to reduce the efficiency of the single quantum
system \cite{6,7,8,9}. In this paper we report the first diamond
based defect centre which behaves as a true two-level system. This
centre also exhibits the highest measured SP emission
rate at room temperature in the MHz regime.\\
\\
Many schemes have been proposed to improve the emission rate from
diamond based SPSs with perhaps the most promising involving the
incorporation of single emitting diamond nano-crystals into cavity
structures \cite{11}. Unfortunately, the lack of mature cavity
technology has limited the incorporation of diamond based SPSs
into such designs \cite{12}. As a result, researchers are in
parallel exploring new defect centres which may result in more
efficient SP emission. For diamond based SPSs to emit more
efficiently new defect centres need to be explored and/or created
which comprise a simple two level energy structure similar to that
observed in semiconductor QDs. To this end, this work explores
novel defect centres which can be created synthetically
using the well known microwave plasma enhanced chemical vapour deposition (MPECVD) technique \cite{13}.\\
\\
The fused silica substrate characterised in this work was
1$\times$1 $\mu$m in size and ultrasonically seeded with
commercial grade diamond powder 0-2 $\mu$m in size prior to MPECVD
growth. The 30 minute diamond growth was conducted under the
conditions described in \cite{13}, with background levels of
nitrogen, nickel and silicon present in the system.
Photo-luminescent studies were conducted on the fused silica
substrate with the aim of identifying novel luminescent defect
centres. Many emitting centres were identified by the in house
fibre based scanning confocal microscope, operating with 532 nm
excitation and a spatial resolution of $\sim$ 400 nm. Figure
\ref{Fig:Confocal images and spectra} (a) illustrates the large
number of emitting centres within a  20$\times$20 $\mu$m area of
the substrate. The photo-luminescence (PL) spectra from each
emitting crystal was measured in the first instance to determine
the type of emitting centre and secondly to identify any
uncharacteristic spectral lines. The room temperature PL spectrum
from the emitting centre identified in the centre of Figure
\ref{Fig:Confocal images and spectra} (b) is shown in Figure
\ref{Fig:Confocal images and spectra} (c).  The spectra reveals a
sharp zero phonon line (ZPL) centered at 734 nm with a full width
at half maximum (FWHM) of 4.1 nm.  The Huang-Rhys factor which is
a measure of the relative intensity of the ZPL compared to the
entire spectrum was found to be 0.81 which is higher than that
reported for the nickel related (NE8) centre \cite{6} and
significantly higher than the well known nitrogen vacancy (N-V)
centre (0.04) \cite{6}.
The photon statistics of the centre were investigated using a
fibre based version of the Hanbury Brown and Twiss interferometer
\cite{19}. The second-order intensity autocorrelation function:\\
\\
$g^{(2)}(\tau)=<\,I(t)I(t+\tau)>/<I(t)>^{2}$\\
\\
obtained from the photon coincidence rate histogram is shown in
Figure \ref{Fig:Anti-bunching vs pump power} for three different
incident pump powers.
The autocorrelation function minimum at the zero delay time,
$g^{(2)}(0)$, was measured to be 0.16, with the deviation from
zero being attributed to the background surrounding and/or within
the emitting crystal. The important conclusion to be drawn from
the autocorrelation measurements, as a function of pump power, is
the absence of `photon bunching' \cite{14,15} in the coincidence
histogram. Photon bunching is observed in all of the single
diamond related defects reported to date \cite{6,7,8} and is
attributed to the presence of a meta-stable quenching/shelving
state in the electronic structure of the atomic system. The
absence of photon bunching in the coincidence histogram for
excitation powers well above saturation is direct evidence of a
two level energy scheme. The power dependence of the second order
correlation function based on a two level energy scheme can be
described by \cite{7}:
\begin{eqnarray}
g^{(2)}(\tau)\simeq 1-\exp ^{(-k_{T}t)} \label{eq:g2as a function
of pump power}
\end{eqnarray}
where $k_{T} = k_{21}+k_{12}$, with $k_{12}$ and $k_{21}$ being
the pumping rate and photon emission rate of the excited state,
respectively.\\
\\
A least squares fit of Eq.(\ref{eq:g2as a function of pump power})
was applied to the measured anti-bunching data as a function of
pump power, as seen in Figure \ref{Fig:Anti-bunching vs pump
power}. The fitting parameter $k_T$ can then be used to obtain the
fluorescence decay rate of the excited state as $k_T(0) = k_{21}$,
in the limit of zero pump excitation $(P\rightarrow0)$. Figure
\ref{Fig:Decay rate vs pump power} shows the fitting parameter
$k_T$ as a function of excitation power.
\begin{figure}[hbt]\centerline{\includegraphics[width=0.8\columnwidth,clip]
{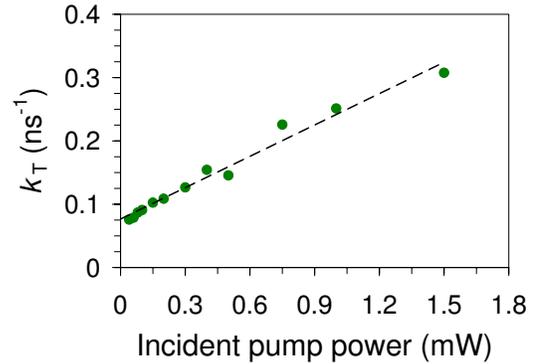}}\caption{Decay rate $k_T$ as
a function of incident pump power. The solid line represents a
linear fit to the data as the fluorescence decay rate is linearly
dependent on the pump excitation.} \label{Fig:Decay rate vs pump
power}
\end{figure}
From the applied linear fit to the data the excited state lifetime
$\tau_2$ = $1/k_{21}$ = 13.6 ns. This lifetime is comparable to
the (N-V) centre in bulk diamond (11 ns) \cite{16} and
considerably longer than those observed from the (NE8) (2 ns)
\cite{14} and silicon-vacancy (Si-V) (1.2 ns) \cite{8} centres;
however due to the two level nature of the atomic centre it is
able to emit more efficiently compared to its three level system
counterparts. This is evidenced by the extraordinary single photon
emission rate (see Figure \ref{Fig:Emission rate vs pump power})
measured as a function of pump power, under 532 nm excitation.
The saturation behaviour of the emission is consistent with that
expected for a single atomic system with a finite lifetime. The
theoretical fit to the measured emission data was derived from the
steady state solution of the excited state population of a two
level system and is given by: $I(P)=I_{sat}P/(P_{sat}+P)$, where
$I_{sat}$ represents the photon saturation count rate, $P$ is the
incident pump power and $P_{sat}$ is the saturation pump power.
From the fit to the experimental data $P_{sat}$ = 224 $\mu$W and
$I_{sat}$ = 1.8 million counts/s. The saturation power is
comparable to that found for N-V centres and considerably less
than that observed for the NE8 and Si-V centres. Moreover, the
saturation count rate of the 734 nm centre is the highest reported
for a single diamond defect and the first observation of stable
MHz emission from a single photon source operating at room
temperature.  The two level nature of the centre also allows an
accurate estimate of the total collection efficiency of the system
to be made as there is no quenching state reducing the number of
emitting photons. Based on the fluorescence lifetime, the centre
emits at 73.5 MHz of which 1.6 MHz was detected; this equates to a
total system collection efficiency of 2.17\%. With further
optimization of the substrate and the potential to manipulate
diamond crystals onto plasmonic structures \cite{17}, there is the
possibility to achieve
SP emission into the tens of MHz range at room temperature.\\
\\
The atomic structure of the single emitting defect centre is at
this stage unclear, there are reports of a 733 nm absorption peak
from Si related centres in single crystal diamond, however no
emission peaks from the 733 nm line have been observed in the PL
under 514 nm excitation \cite{18}. One can speculate though that
the origin of the centre lies within a mixture of the constituents
present in the diamond reactor during growth which includes
nitrogen, silicon, nickel and hydrogen. PL studies of the diamond
seed material before growth under 532 nm excitation revealed N-V
centres were present in a number of seed crystals, however, this
fluorescence emitter was only observed after being exposed to the
MPECVD process.\\
\\
A new and highly efficient diamond-based single photon source
operating in the infra red at 734 nm has been fabricated and
characterised. The diamond defect exhibits an intense and
spectrally narrow emission line and is the first SPS to exhibit
MHz operation at room temperature. The centre is found to
significantly extend the performance range of diamond based SPSs
due to its two level nature. With the opportunity to manipulate
such sources onto structures, which can further enhance the
emission properties, the future for practical room temperature
quantum devices looks bright.
\begin{acknowledgments}
The authors gratefully acknowledge Igor Aharonovich and Stefania
Castelletto for many helpful discussions. This project was
supported by Quantum Communications Victoria, which is funded by
the Victorian Government's Science, Technology and Innovation
initiative.
\end{acknowledgments}

\end{document}